\documentclass[conference,compsoc]{IEEEtran}

\usepackage[utf8]{inputenc}
\usepackage[T1]{fontenc}

\usepackage{balance}
\usepackage{tikz}
\usepackage{standalone}
\usetikzlibrary{shapes}
\usetikzlibrary{trees}

\usepackage{algorithmic}
\usepackage{algorithm}
\usepackage{url}
\usepackage{graphicx}
\usepackage{color}
\usepackage{listings}
\usepackage{subfigure}
\usepackage{xspace}
\usepackage[binary-units]{siunitx}
\usepackage{booktabs}
\usepackage{multirow}
\usepackage{etoolbox}

\usepackage{amsmath}
\usepackage{amssymb}

\newcommand{\mpineighborallgather}{\texttt{MPI\_\-Neighbor\_\-allgather}\xspace}
\newcommand{\mpineighboralltoall}{\texttt{MPI\_\-Neighbor\_\-alltoall}\xspace}

\newcommand{\mpineighboralltoallw}{\texttt{MPI\_\-Neighbor\_\-alltoallw}\xspace}

\newcommand{\isoneighboralltoallinit}{\texttt{Iso\_\-neighbor\_\-alltoall\_\-init}\xspace}
\newcommand{\isoneighboralltoallwinit}{\texttt{Iso\_\-neighbor\_\-alltoallw\_\-init}\xspace}
\newcommand{\isoneighboralltoallvinit}{\texttt{Iso\_\-neighbor\_\-alltoallv\_\-init}\xspace}

\newcommand{\isoneighboralltoall}{\texttt{Iso\_\-neighbor\_\-alltoall}\xspace}
\newcommand{\isoneighboralltoalldirect}{\texttt{Iso\_\-neighbor\_\-alltoall\_\-direct}\xspace}
\newcommand{\isoneighboralltoallw}{\texttt{Iso\_\-neighbor\_\-alltoallw}\xspace}
\newcommand{\isoneighborallgather}{\texttt{Iso\_\-neighbor\_\-allgather}\xspace}
\newcommand{\isostart}{\texttt{Iso\_\-Start}\xspace}

\newcommand{\mpistruct}{\texttt{MPI\_\-Type\_\-create\_\-struct}\xspace}

\newcommand{\mpidimscreate}{\texttt{MPI\_\-Dims\_\-create}\xspace}

\newcommand{\mpidistgraphcreate}{\texttt{MPI\_\-Dist\_\-graph\_\-create}\xspace}
\newcommand{\mpidistgraphcreateadjacent}{\texttt{MPI\_\-Dist\_\-graph\_\-create\_\-adjacent}\xspace}

\newcommand{\isoneighborhoodcreate}{\texttt{Iso\_\-neighborhood\_\-create}\xspace}

\newtheorem{proposition}{Proposition}

\newcommand{\fig}{Figure\xspace}

\newcommand{\tab}{Table\xspace}
\newcommand{\alg}{Algorithm\xspace}

\newcommand{\Sec}{Section\xspace}

\newcommand{\lst}{Listing\xspace}

\newcommand{\alltoall}{all-to-all\xspace}
\newcommand{\alltoallw}{all-to-allw\xspace}
\newcommand{\allgather}{allgather\xspace}
\newcommand{\Alltoall}{All-to-all\xspace}

\newcommand{\Allgather}{Allgather\xspace}

\newcommand{\ie}{i.e.\@\xspace}
\newcommand{\etal}{\textit{et~al.\@}\xspace}

\newcommand{\jupiternecmpi}{NEC\,MPI~1.3.1\xspace}

\newcommand{\mpiruns}{\texttt{mpirun}'s\xspace}
\newcommand{\archermpi}{Cray\,MPICH~7.2.6\xspace}

\newcommand{\vscmvapich}{MVAPICH~2-2.2b}

\newcommand{\machjupiter}{Jupiter\xspace}
\newcommand{\machvsc}{\mbox{VSC-3}\xspace}
\newcommand{\macharcher}{ARCHER\xspace}

\newcommand{\infiniband}{InfiniBand\xspace}

\newcommand{\runtime}{run-time\xspace}
\newcommand{\runtimes}{run-times\xspace}

\newcommand{\Moore}{Moore\xspace}

\newcommand{\setup}{set-up\xspace}
\newcommand{\Setup}{Set-up\xspace}

\makeatletter
\AtBeginEnvironment{lstlisting}{%
\long\def\@makecaption#1#2{%
\@IEEEfigurecaptionsepspace
\setbox\@tempboxa\hbox{\normalfont\footnotesize {#1.}\nobreakspace\nobreakspace #2}%
\ifdim \wd\@tempboxa >\hsize%
\setbox\@tempboxa\hbox{\normalfont\footnotesize {#1.}\nobreakspace\nobreakspace}%
\parbox[b]{\hsize}{\normalfont\footnotesize\noindent\unhbox\@tempboxa#2}\medskip%
\else%
\ifCLASSOPTIONconference \hbox to\hsize{\normalfont\footnotesize\hfil\box\@tempboxa\hfil}%
\else \hbox to\hsize{\normalfont\footnotesize\box\@tempboxa\hfil}\medskip%
\fi\fi }}
\makeatother

\newcommand{\even}{\mathrm{even}}

\newcommand{\remind}[1]{}

\newcommand{\leaveout}[1]{}

\begin{document}
\IEEEoverridecommandlockouts
\title{Message-Combining Algorithms for Isomorphic, Sparse Collective
  Communication\thanks{This work was first submitted for double blind review 
on the 28th of January, and again on the 8th of April, 2016.}}

\author{%
\IEEEauthorblockN{Jesper Larsson Tr\"aff,
Alexandra Carpen-Amarie,
Sascha Hunold,
Antoine Rougier}
\IEEEauthorblockA{TU Wien\\
Faculty of Informatics, Institute of Information Systems\\
Research Group Parallel Computing\\
Favoritenstrasse 16/184-5, Vienna, Austria\\ 
Email: \url{{traff,carpenamarie,hunold,rougier}@par.tuwien.ac.at}}%
}

\maketitle

\begin{abstract}
Isomorphic (sparse) collective communication is a form of collective
communication in which all involved processes communicate in small,
identically structured neighborhoods of other processes.  Isomorphic
neighborhoods are defined via an embedding of the processes in a
regularly structured topology, e.g., $d$-dimensional torus, which may
correspond to the physical communication network of the
underlying system. Isomorphic collective communication is useful for
implementing stencil and other regular, sparse distributed
computations, where the assumption that all processes behave (almost)
symmetrically is justified.

In this paper, we show how efficient message-combining communication
schedules for isomorphic, sparse collective communication can easily
and efficiently be computed by purely local computations. We give
schemes for \emph{isomorphic \alltoall} and \emph{\allgather}
communication that reduce the number of communication rounds and
thereby the communication latency from $s$ to at most $Nd$, for
neighborhoods consisting of $s$ processes with the (small) factor $N$
depending on the structure of the neighborhood and the capabilities of
the communication system. Using these schedules, we give
\emph{zero-copy implementations} of the isomorphic collectives using
MPI and its derived datatypes to eliminate explicit, process-local
copy operations. By benchmarking the collective communication
algorithms against straightforward implementations and against the
corresponding MPI neighborhood collectives, we document significant
latency improvements of our implementations for 
block sizes of up to a few kilobytes. We discuss further optimizations
for computing even better schedules, some of which have been
implemented and benchmarked.

The proposed message-combining schedules are difficult to incorporate
as implementations for the MPI neighborhood collectives, since MPI
processes lack the necessary information about the global
structure of the communication pattern.  If this information is
externally asserted, our algorithms can be used to improve the
performance of the MPI neighborhood collectives for isomorphic
communication patterns for latency-sensitive problem sizes.
\end{abstract}

\section{Introduction}
\label{sec:introduction}

Structured, sparse communication patterns appear frequently in
parallel numerical applications, notably stencil-patterns in two and
higher dimensions~\cite{Sourcebook03,Epperson07,CuiOlsen10}. With MPI
3.0 and later versions of the \emph{Message-Passing
  Interface}~\cite{MPI-3.1}, sparse communication patterns can be
expressed as so-called \emph{neighborhood collective operations}. The
specific mechanism of MPI relies on virtual process topologies to
define communication neighborhoods for the ensuing neighborhood
collective operations. In many respects this is undesirable. The
neighborhood that is implicit with Cartesian communicators is the set
of immediate distance one neighbors along the dimensions, thus
collective communication in a standard, 2-dimensional, 9-point (and
3-dimensional, 27-point, etc.) stencil pattern cannot be expressed
with Cartesian communicators. The general, distributed graph topology
interface allows specification of arbitrary, directed communication
graphs, and can thus express any desired stencil communication
pattern. However, information about the global, highly regular
structure of the communication graph is not conveyed to the MPI
library, which makes many types of beneficial optimizations difficult
and/or computationally hard.

We address these problems, and examine a restricted form of MPI-like,
sparse, collective communication which we term \emph{isomorphic,
  sparse collective communication}~\cite{Traff15:isosparse}.
Isomorphic, sparse collective communication means that all processes
communicate in structurally similar patterns and that this property is
asserted to the processes. Concretely, the MPI processes are assumed
to be placed in some regular (virtual) topology, like for instance a
$d$-dimensional torus. A sparse process neighborhood is described by a
list of relative, $d$-dimensional vector offsets. In this situation,
process neighborhoods are isomorphic if the offset lists are identical
(same vector offsets in the same order) over all processes.  The
proposed interfaces are persistent both in the sense that the same
sparse, isomorphic neighborhood can be used in different communication
operations, and that operations with the same buffer and datatype
parameters can be performed several times. The persistent interfaces
provide handles to precompute communication schedules such that the
costs of the schedule computation can be amortized over several,
actual collective communication operations. For the isomorphic
collective operations discussed here, the schedule computation is
actually very fast, but the setting up of the MPI derived datatypes
(that are used to make data blocks move between intermediate and final
result buffers) consumes enough time to make persistence worthwhile.

The main contribution of this paper is to show that efficient,
deadlock-free, message-combining communication schedules for
\emph{isomorphic \alltoall} and \emph{\allgather} can be easily
computed, given the isomorphic assertion that all processes use the
exact same, relative neighborhood. The resulting message-combining
schedules correspond to communication optimizations typically made for
$9$- and $27$-point stencils in two and three dimensions,
respectively, where messages to corner processes piggyback on messages
sent along the principal dimensions. 
However, our algorithms are general, and work for any isomorphic
neighborhood, such that also asymmetric patterns can be catered
to. For sparse neighborhoods consisting of $s$ neighbors,
message-combining reduces the number of communication rounds from $s$
send and receive rounds of a straightforward, linear algorithm to
$Nd$, where the constant $N$ depends on the structure of the
neighborhood and on assumptions about the underlying communication
system. For instance, the number of rounds in a $27$-point stencil
pattern in a $3$-dimensional mesh or torus is reduced from $26$ to
only $6$, under the assumption of a one-ported communication
system. This is achieved by combining messages to different neighbors
and sending larger, combined messages along the torus dimensions
only. Since some messages are thus sent via several, intermediate
processes, there is often a tradeoff between number of rounds and
total communication volume, as is the case for dense \alltoall
communication~\cite{Bruck97}. Message-combining is implemented using
the MPI derived datatype mechanism to specify for each communication
round which messages have to be sent and received and from which
communication buffers. Allowing space for an intermediate
communication buffer, a per-message double-buffering scheme can be
implemented in this manner, thereby completely eliminating explicit
message copying or packing/unpacking and leading to our resulting
\emph{zero-copy implementations}. We have used similar techniques
previously in~\cite{Traff14:bruck}.

Our first \alltoall and \allgather algorithms assume a one-ported
torus communication network, and are round- and volume-optimal under
this assumption.  We have implemented these algorithms, both in
regular and irregular versions, and present an extensive benchmark
evaluation with comparisons to both the current MPI 3.1 neighborhood
collective implementations and the straightforward, $s$-communication
round implementations of the isomorphic interfaces. For small message
sizes up to a few kilobytes, the experimental results show the
expected reduction in communication time. Furthermore, for larger
neighborhoods in three and higher dimensions, we observe very
substantial improvements.

For our second set of algorithms we relieve the restriction of only
immediate torus neighbor communication, and allow direct communication
along the torus dimensions. For neighborhoods with long-distance
neighbors, this can lead to significant reductions in the number of
communication rounds, which now depends only on the number of
different coordinate values in each dimension, and not on the
magnitude of the coordinates. A second set of experiments illustrates
the effects of the fewer communication rounds.  Further relieving
network assumptions leads to interesting optimization problems for
minimizing the number of communication rounds or maximizing the number
of ports that can be used per communication round. We discuss some of
these problems.
 
There is a large amount of work on optimizations for stencil
computations,
see~\cite{BasuHallWilliamsStraalenOlikerColella15,Dursun09,Dursun12,StengelTreibigHagerWellein15,TangChowdhuryKuzmaulLukLeiserson11}
for some that has influenced this work, many of which also discuss
communication
optimizations~\cite{BordawekarChoudharyRamanujam96:automatic}.
Stencil computations have been used to analyze (implications of) new
MPI one-sided communication support by
Zhu~\etal~\cite{ZhuZhangYoshiiLiZhangBalaji15}.  General optimization
techniques for the MPI neighborhood collectives were proposed by
Hoefler and Schneider~\cite{HoeflerSchneider12}, who do not exploit
external assertions about the overall structure of neighborhoods to
simplify, e.g., scheduling by coloring.  More general, dynamic
neighborhood communication on top of MPI is discussed by
Ovcharenko~\etal~\cite{Ovcharenko12}.  Souravlas and
Roumeliotis~\cite{SouravlasRoumeliotis08:torus} also considered
message-combining optimizations but in a more limited context than done
here.

\section{Isomorphic, Sparse Collective Communication}
\label{sec:definitions}

We now describe more formally what is meant by isomorphic, sparse
collective communication. The notation introduced here will be used
for the remainder of the paper. We show the concrete interfaces as
implemented in our library.

An isomorphic, sparse collective communication pattern is defined
relative to some given, structured organization of the processes.  Let
$p$ be the number of processes, and assume that they are organized in
a $d$-dimensional torus with dimension sizes $p_0, p_1,\ldots,p_{d-1}$
and $\Pi_{i=0}^{d-1}p_i=p$. Each ranked process~$R, 0\leq R<p$ is
identified by a coordinate $(r_0,r_1,\ldots r_{d-1})$ with $0\leq
r_i<p_i$ for $i=0,\ldots, d-1$.

A (sparse) \emph{$s$-neighborhood} of a process is a collection of
$s$ processes to which the process shall \emph{send} data. The
collection is given as a sequence of $s$ \emph{relative-coordinate
  vectors} $\allowbreak C^0, \allowbreak C^1,\ldots \allowbreak
C^{s-1}$. Each $C^i$ has the form $(c^i_0, \allowbreak c^i_1, \ldots,
\allowbreak c^i_{d-1})$ for arbitrary integer offsets $c^i_j$
(positive or negative). A set of identical $s$-neighborhoods for a set
of processes is said to be \emph{isomorphic}. An \emph{isomorphic,
  sparse collective operation} is a collective operation over $p$
processes with isomorphic neighborhoods. Note that an $s$-neighborhood
is allowed to have repetitions of relative coordinates, and that a
process can be a neighbor of itself, for instance if relative coordinate
$(0,0,\ldots,0)$ is in the $s$-neighborhood. Also note that different
coordinates may denote the same neighbor, which can easily happen if
$p$ is small.

We define torus vector addition $\oplus$ for vectors $R$ and
$C$ in the given torus by $R\oplus C = ((r_0+c_0)\bmod p_0,
(r_1+c_1)\bmod p_1, \ldots, (r_{d-1}+c_{d-1})\bmod p_{d-1})$.  Each
process $R=(r_0, r_1,\ldots, r_{d-1})$ with $s$-neighborhood
$\allowbreak C^0, \allowbreak C^1, \ldots, \allowbreak C^{s-1}$ shall
send data to the $s$ \emph{target processes} $R\oplus C^i$ for
$i=0,\ldots, s-1$.  Since neighborhoods are isomorphic, it follows
that the process will need to receive data from $s$ \emph{source
  processes} $R\ominus C^i$.

The concrete, isomorphic, sparse collective operations we consider
here are of the \alltoall and the \allgather type. In an
\emph{isomorphic \alltoall communication}, each process sends an
individual, possibly different \emph{block} of data to each of its
target neighbors, and receives a block of data from each of its source
neighbors. In an \emph{isomorphic \allgather communication}, each
process sends the \emph{same block} of data to each of its target
neighbors, and receives a block of data from each of its corresponding
sources.

\begin{lstlisting}[float,caption={
The collective, isomorphic neighborhood \setup function. Calling
processes must supply the same list of relative coordinates.
},
label=lst:isocreate,
floatplacement=ht!
]
Iso_neighborhood_create(MPI_Comm cartcomm, 
    int s, int relative_coordinates[], 
    MPI_Comm *isocomm)
\end{lstlisting}

\begin{lstlisting}[float,caption={
 The interfaces for regular, persistent, isomorphic \alltoall
  and \allgather communication. Collective communication is initiated
  and completed by the start call, which uses the buffer and datatype
  parameters given in the corresponding init call. The neighborhood is
  defined by the isomorphic communicator created by a previous \setup
  call, and send and receive buffers must be large enough to store the
  data blocks sent to and received from the neighbors.
},
label=lst:isoall,
floatplacement=ht!
]
Iso_neighbor_alltoall_init(void *sendbuf, 
    int sendcount, MPI_Datatype sendtype,
    void *recvbuf,
    int recvcount, MPI_Datatype recvtype,
    MPI_Comm isocomm, Iso_request *request)

Iso_neighbor_allgather_init(void *sendbuf, 
    int sendcount, MPI_Datatype sendtype,
    void *recvbuf, 
    int recvcount, MPI_Datatype recvtype,
    MPI_Comm isocomm, Iso_request *request)

Iso_start(Iso_request *request);

Iso_request_free(Iso_request *request);
\end{lstlisting}

For a library on top of MPI, the corresponding interface functions are
as follows. First, the MPI processes need to be organized in a
$d$-dimensional Cartesian mesh or torus with a suitable
$d$-dimensional Cartesian communicator
(\texttt{cartcomm})~\cite[Chapter 7]{MPI-3.1}.  The isomorphic
neighborhood \setup function is called on this communicator, and takes
a list of neighbor coordinates given as a one-dimensional, flattened
array of relative coordinates, and attaches this to a new communicator
\texttt{isocomm}. The \setup operation is collective, and a strict
requirement is that the calling processes all give the \emph{exact
  same} list of relative neighbor coordinates. The function prototype
is shown in \lst~\ref{lst:isocreate}. As an example, assume we want to
perform isomorphic \alltoall to the processes in the positive octant
of a three-dimensional torus. The relative coordinates are
$(1,0,0),\allowbreak (0,1,0),\allowbreak (0,0,1),\allowbreak (1,1,0),
\allowbreak (1,0,1),\allowbreak (0,1,1),\allowbreak (1,1,1)$ (and
$(0,0,0)$ if the process has a message to itself). The corresponding
call would be
\begin{lstlisting}
int octant[] = 
{1,0,0,0,1,0,0,0,1,1,1,0,1,0,1,0,1,1,1,1,1};
Iso_neighborhood_create(cartcomm,
                        7,octant,&isocomm);
\end{lstlisting}
Any permutation of the 7 neighbors would specify the same neighborhood
(provided that all calling processes give the neighbors in the same
order; if not, the outcome of the call and of ensuing communication
operations is undefined and either may deadlock), but the
order is important and determines the order of the message blocks in
the send and receive buffers of the isomorphic communication operations.

\begin{lstlisting}[float,caption={
 The interfaces for irregular, persistent, isomorphic \alltoall
  and \allgather communication. 
},
label=lst:isoirreg,
floatplacement=ht!
]
Iso_neighbor_alltoallv_init(void *sendbuf, 
    int sendcounts[], MPI_Aint senddispls[], 
    MPI_Datatype sendtype,
    void *recvbuf,
    int recvcounts[], MPI_Aint recvdispls[], 
    MPI_Datatype recvtype,
    MPI_Comm isocomm, Iso_request *request)

Iso_neighbor_allgatherv_init(void *sendbuf, 
    int sendcount, MPI_Datatype sendtype,
    void *recvbuf, 
    int recvcounts[], MPI_Aint recvdispls[], 
    MPI_Datatype recvtypes,
    MPI_Comm isocomm, Iso_request *request)

Iso_neighbor_alltoallw_init(void *sendbuf, 
    int sendcounts[], MPI_Aint senddispls[], 
    MPI_Datatype sendtypes[],
    void *recvbuf,
    int recvcounts[], MPI_Aint recvdispls[], 
    MPI_Datatype recvtypes[],
    MPI_Comm isocomm, Iso_request *request)

Iso_neighbor_allgatherw_init(void *sendbuf, 
    int sendcount, MPI_Datatype sendtype,
    void *recvbuf, 
    int recvcounts[], MPI_Aint recvdispls[], 
    MPI_Datatype recvtypes[],
    MPI_Comm isocomm, Iso_request *request)
\end{lstlisting}
The collective interface consists in two parts, namely an init call
where a communication schedule can be precomputed, and an ensuing
communication start call. This separation allows the reuse of a
communication schedule computed in the init call over a number of
collective communication operations with the same buffer and datatype
parameters.  The idea is similar to the persistent point-to-point
communication operations of MPI~\cite[Section
  3.9]{MPI-3.1}\footnote{There is so far no persistent collectives
  counterpart in MPI. This is being considered by the MPI Forum.}. The
interface functions that we have implemented are shown in
\lst~\ref{lst:isoall}, and have the usual MPI flavor. The data blocks
for the target neighbors are stored consecutively at the
\texttt{sendbuf} address in the order determined by the order of the
neighbors; similarly, blocks from the source neighbors will be stored
at the \texttt{recvbuf} address in the same order. In the regular
variants of the isomorphic collectives, all blocks have the same size
and structure as determined by the count and MPI datatype
arguments. Irregular \alltoall versions, \ie,
\isoneighboralltoallwinit and \isoneighboralltoallvinit, are defined
analogously, and are shown in \lst~\ref{lst:isoirreg}. The requirement
for these irregular versions is that all processes specify exactly the
same block sizes via count and datatype arguments, and that send and
receive block sizes match pairwise. Note that the isomorphic
requirement in neither regular nor irregular case means that processes
have to use the same datatype arguments; also the datatype for the
receive and the send buffers may be different. The regular variants of
the collectives only require that blocks all have the same size, whereas the
irregular variants require blocksizes to be pairwise equal.

\section{Message-Combining Algorithms}

\begin{lstlisting}[float,caption={
Straightforward, isomorphic \alltoall communication in $s$ communication
  rounds.  The implementation is deadlock free, since all processes have
  specified the neighborhood by identical lists of relative
  coordinates.
  },
label=lst:s-round,
floatplacement=t
]
// R: process rank as d-dimensional vector
// C[i]: i-th offset vector from isocomm
// rank(C): linear MPI rank of vector C
for (i=0; i<s; i++)
  MPI_Sendrecv(sendbuf[i],...,rank(R+C[i]),
               recvbuf[i],...,rank(R-C[i]),
               isocomm); 
\end{lstlisting}

We now show how the isomorphic neighborhood assertion makes it easy to
precompute good, message-combining communication schedules.  First
note that the simple scheme in \lst~\ref{lst:s-round} is correct
(deadlock free). Each process looks up its rank as a $d$-dimensional
vector $R$ in the underlying torus, and uses the coordinate offsets to
compute source and target ranks as explained in the previous section.
In the $i$th of $s$ communication rounds, it sends and receives blocks
directly to and from the $i$th source and target processes. Although
the algorithm is trivial, it is worth pointing out that deadlock
freedom follows from the assumption that neighborhoods are isomorphic. In
round $i$ when process $R$ is sending block $i$ to target neighbor
$R\oplus C^i$, this neighbor expects to receive a block from its $i$th
source process, which is indeed $(R\oplus C^i)\ominus C^i=R$.  For
neighborhoods defined by unrestricted communication graphs as it is
the case with MPI distributed graph communicators, or if the processes
had given their list of neighbors in different orders, this would not
be the case, and the scheme can deadlock.

The $s$-round algorithm assumes that messages can be sent directly
from a process to its target neighbors, and performs one send and
receive operation per communication round. It can trivially be
extended to exploit $k$-ported communication systems also for $k>1$ by
sending and receiving instead $k$ blocks per round.  Our first goal is
to provide message-combining schemes with fewer communication rounds,
and to precompute schedules that for each process tell which
(combined) message blocks to send and receive in each communication
round. Our schedules will have the property that all processes follow
the same steps, and can be computed locally for each process from its
list of neighbors.

For the algorithm design, we first assume that the underlying
communication network is a bidirectional (send-receive), one-ported,
$d$-dimensional torus, such that communication is allowed only along
the $d$ dimensions, and only between immediate neighbors. Only one
dimension can be actively communicating at any one instant, but a
process can simultaneously send and receive a message in the given
dimension.  We stress that the torus assumption is made to help the
algorithm design, and is not necessarily an assumption about the
underlying hardware. The dimensions are processed in some order, and
in each iteration all blocks that have to go along one dimension are
sent together as one message. This reduces the number of communication
operations (and start-up latencies) from $s$ to $O(d)$. The schedules
for \alltoall and \allgather communication operations are explained
and analyzed in more detail below.  The key observation is that
schedules can be developed from the processes point of view by
analyzing the $s$-neighborhood of relative coordinates. As in
\lst~\ref{lst:s-round}, processes will follow the same schedule from
which deadlock freedom and correctness follow.  In each communication
round, all processes will have the same (relative) blocks to forward
to other processes.  Blocks are always routed along shortest paths in
the torus network, but may pass through processes that are not in the
neighborhood.

\subsection{\Alltoall Schedule}

Define the norm of vector $C=(c_0, \allowbreak c_1, \ldots,
\allowbreak c_{d-1})$ by $\|C\|=\sum_{j=0}^{d-1}|c_j|$.  This norm
counts how many communication steps are needed in the torus to route a
block from (any) process $R$ to its target neighbor $R\oplus C$. The
block can be (minimally) routed from $R$ to $R\oplus C$ by sending it
successively $c_j$ (positive or negative) hops along dimension $j$ for
$j=0,\ldots, d-1$.  All $s$ blocks from process $R$ to its relative
neighbors $C^i$ are routed as follows in $d$ rounds.  In round $j$
each process will be handling blocks to be passed along dimension
$j$. To route all blocks along dimension $j$,
$\max_{i=0}^{s-1}(\max(c^i_j,0)+\max(-c^i_j,0))$ communication steps
are necessary. In step $h$, for each coordinate $|c^i_j|>h$, an old
block is sent and a new one received, with all such blocks combined
into a single message. By the end of a communication round, all blocks
of a process will have been routed the corresponding $c^i_j$ hops
ahead, and after all $d$ rounds, all blocks have been received by
their target processes.

Since a process can, in each communication step, only send and receive
along one dimension, it follows that in total $D =
\sum_{j=0}^{d-1}\max_{i=0}^{s-1}(\max(c^i_j,0)+\max(-c^i_j,0))$
communication steps are required, which is exactly the number of steps
performed by the algorithm. Since communication is only done between
direct torus neighbors, the shortest path for each block is $\|C^i\|$
hops, such that the total number of blocks sent per process
(communication volume) is $V=\sum_{i=0}^{s-1}\|C^i\|$. Also this is
achieved by the algorithm.  For a given (isomorphic) $s$-neighborhood,
both $D$ and $V$ can be easily computed and used to estimate the cost
of the \alltoall communication. In a simple, linear cost model with
latency $\alpha$ and cost per unit $\beta$, this would be $D\alpha +
\beta Vm$ for blocks of $m$ units. In this cost model, the
message-combining schedule can be faster than the direct schedule for
fully connected, bidirectional networks of \lst~\ref{lst:s-round}, if
$D\alpha + \beta Vm < s(\alpha+\beta m)$, that is
$m<\frac{\alpha}{\beta}\frac{s-D}{V-s}$ for $s<V$ and $D<s$.
We have argued for the following statement.

\begin{proposition}
\label{prop:alltoall}
In $d$-dimensional, 1-ported, bidirectional tori, isomorphic \alltoall
communication in $s$-neighborhoods with blocks of size $m$ can be
performed round- and volume-optimally in $D$ communication rounds and
total communication volume $Vm$. A corresponding schedule can be
computed in $O(sD)$ operations.
\end{proposition}

The schedule computation is described in detail in
\Sec~\ref{sec:zerocopy} from which the stated bound follows.  If
coordinates of all $s$ neighbors are bounded (each
$c_j=0,1,2,3,\ldots,k$ for some small constant $k$), then $D\leq kd$,
and the number of communication rounds will be small.

\subsection{\Allgather Schedule}

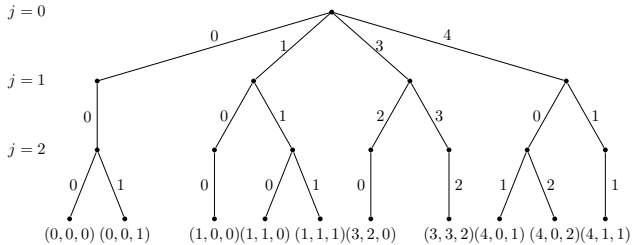
\begin{figure}
\centering
        \begingroup
        \tikzstyle{end} = [circle, minimum width=3pt,fill, inner sep=0pt]
\begin{tikzpicture}[grow=down]
	\tikzstyle{level 1}=[sibling distance=34mm]
	\tikzstyle{level 2}=[sibling distance=17mm]
	\tikzstyle{level 3}=[sibling distance=12mm]

	\node[end] {}
	coordinate(zero)
	child {
		coordinate(one)
		node[end]{}
		child {
			coordinate(two)
			node[end]{}        
			child {
				node[end, label=below:
				{$(0,0,0)$}] {}
				edge from parent
				node[below,left] {$0$}
			}
			child {
				node[end, label=below:
				{$(0,0,1)$}] {}
				edge from parent
				node[below,right] {$1$}
			}
			edge from parent 
			node[below,left] {$0$}
		}
		edge from parent 
		node[below,above] {$0$}
	}
	child {
		node[end] {}
		child {
			node[end] {}
			child {
				node[end, label=below:
				{$(1,0,0)$}] {}
				edge from parent
				node[below,left] {$0$}
			}
			edge from parent
			node[below,left] {$0$}
		}
		child {
			node[end] {}
			child {
				node[end, label=below:
				{$(1,1,0)$}] {}
				edge from parent
				node[below,left] {$0$}
			}
			child {
				node[end, label=below:
				{$(1,1,1)$}] {}
				edge from parent
				node[below,right] {$1$}
			}
			edge from parent
			node[below,right] {$1$}
		}
		edge from parent 
		node[below,left] {$1$}
	}
	child {
		node[end] {}
		child {
			node[end] {}
			child {
				node[end, label=below:
				{$(3,2,0)$}] {}
				edge from parent
				node[below,left] {$0$}
			}
			edge from parent
			node[below,left] {$2$}
		}
		child {
			node[end] {}
			child {
				node[end, label=below:
				{$(3,3,2)$}] {}
				edge from parent
				node[below,right] {$2$}
			}
			edge from parent
			node[below,right] {$3$}
		}
		edge from parent 
		node[below,right] {$3$}
	}
	child {
		node[end] {}
		child {
			node[end] {}
			child {
				node[end, label=below:
				{$(4,0,1)$}] {}
				edge from parent
				node[below,left] {$1$}
			}
			child {
				node[end, label=below:
				{$(4,0,2)$}] {}
				edge from parent
				node[below,right] {$2$}
			}
			edge from parent
			node[below,left] {$0$}
		}
		child {
			node[end] {}
			child {
				node[end, label=below:
				{$(4,1,1)$}] {}
				edge from parent
				node[below,right] {$1$}
			}
			edge from parent
			node[below,right] {$1$}
		}
		edge from parent 
		node[below,above] {$4$}
	};

	\draw[] ([xshift=-61mm]zero) node[left] {$j=0$};
	\draw[] ([xshift=-10mm]one) node[left] {$j=1$};
	\draw[] ([xshift=-10mm]two) node[left] {$j=2$};
\end{tikzpicture}%
        \endgroup
    
\caption{A prefix trie for some $s$-neighborhood. Weights on edges
  from dimension level $j$ to dimension level $j+1$ correspond to the
  $j$th coordinate of some neighbor in the $s$-neighborhood. Each node
  at level $j$ represents the neighbors that share a common prefix of
  $j-1$ coordinates.}
\label{fig:prefixtrie}
\end{figure}

We also use dimension-wise routing for the isomorphic \allgather
operation.  The observation here is that for all relative neighbors
$C^i$ that share a common prefix, i.e., have the same first $j$
coordinates for some $j<d$, the block has to be routed only once to
that prefix (recall that the \allgather operation communicates the
same block to all target neighbors). We construct a prefix-trie as
illustrated in \fig~\ref{fig:prefixtrie}.  In order to ensure that
prefixes get as long as possible, we assume that the order in which
coordinates are visited is in decreasing number of neighbors having
the same coordinate value. Starting from dimension \num{0} there is an
outgoing edge to dimension \num{1} for each different coordinate at
index \num{0} in the $s$-neighborhood. From a node at dimension $j$
with prefix $P_j$ (corresponding to the path from the root of the
trie), representing a block that has been received at that point from
process $R\ominus P_j$, there are outgoing edges to dimension $j+1$
for each different coordinate at index $j$ of the relative neighbors
$C^i$ sharing prefix $P_j$. The leaf nodes of the trie represent the
blocks that will have been received after the $d$ rounds. The prefixes
corresponding to the nodes in the trie can be found by sorting the $s$
neighbor vectors lexicographically. When routing in round $j$, the
number of nodes at level $j$ is the number of different blocks that
have to be sent in that round, and the edges determine the number of
hops that each of these blocks have to be sent.  As in the \alltoall
schedule, in each round~$j$,
$\max_{i=0}^{s-1}(\max(c^i_j,0)+\max(-c^i_j,0))$ communication steps
are necessary, but the communication volume is smaller. The number of
different blocks $W$ received per process throughout the algorithm is
the sum of all weighted path lengths in the trie from the root to the
leaves. Each coordinate value associated with a trie edge determines
the number of hops a certain block is sent in some round.  Note that
$W\leq V$, so for each fixed $s$-neighborhood, \allgather is
potentially less costly than \alltoall. Lexicographic sorting can be
done by bucket sort in $O(sd)$ operations. We have argued informally
for the following statement.

\begin{proposition}
In $d$-dimensional, 1-ported, bidirectional tori, isomorphic \allgather
communication in $s$-neighborhoods with blocks of size $m$ can be
performed round- and volume-optimally in $D$ communication rounds and a total
communication volume $Wm$. A schedule can be computed in $O(sD)$ operations.
\end{proposition}

By the same argument as for \alltoall, $D$ communication rounds are
necessary: there is some neighbor with $j$th coordinate
$\max(c^i_j,0)$ and some with $\max(-c^i_j,0)$, and since
communication is one-ported, this many steps are required for each of
the $j$ rounds.

\subsection{Zero-Copy Implementations}
\label{sec:zerocopy}

\begin{algorithm}[h!]
\begin{algorithmic}
\STATE $\forall 0\leq i<s: \mathrm{hops}[i] \leftarrow \|C^i\|$
\STATE $D\leftarrow 0; V\leftarrow 0$ 
\FOR{$j\leftarrow 0,\ldots,d-1$} 
\FOR{$h\leftarrow 0,\ldots,(\max_{i=0}^{s-1}c_j^i)-1$}
\STATE $k\leftarrow 0; D\leftarrow D+1$ // Positive coordinates
\FOR{$i\leftarrow 0,\ldots,s-1$} 
\IF{$h<c^i_j$}
\IF{$\textrm{firstround}[i]\equiv j$}
\IF{$\even(\mathrm{hops}[i])$}
\STATE $\mathrm{RECV}(\mathrm{part}\ k)\leftarrow \mathtt{interbuf}[i]$ 
\STATE $\mathrm{SEND}(\mathrm{part}\ k)\leftarrow \mathtt{sendbuf}[i]$
\ELSE
\STATE $\mathrm{RECV}(\mathrm{part}\ k)\leftarrow \mathtt{recvbuf}[i]$ 
\STATE $\mathrm{SEND}(\mathrm{part}\ k)\leftarrow \mathtt{sendbuf}[i]$
\ENDIF
\ELSE
\IF{$\even(\mathrm{hops}[i])$}
\STATE $\mathrm{RECV}(\mathrm{part}\ k)\leftarrow \mathtt{interbuf}[i]$ 
\STATE $\mathrm{SEND}(\mathrm{part}\ k)\leftarrow \mathtt{recvbuf}[i]$
\ELSE
\STATE $\mathrm{RECV}(\mathrm{part}\ k)\leftarrow \mathtt{recvbuf}[i]$ 
\STATE $\mathrm{SEND}(\mathrm{part}\ k)\leftarrow \mathtt{interbuf}[i]$
\ENDIF
\ENDIF
\STATE{$V\leftarrow V+1;\mathrm{hops}[i] \leftarrow \textrm{hops}[i]-1; k\leftarrow k+1$}
\ENDIF
\ENDFOR
\ENDFOR
\FOR{$h\leftarrow 0,\ldots,(\max_{i=0}^{s-1}-c_j^i)-1$} 
\STATE // Negative coordinates (analogous)
\ENDFOR
\ENDFOR
\end{algorithmic}
\caption{Computing the alternating, zero-copy \alltoall schedule for
  $s$ blocks and neighbors in $O(sD)$ operations. User buffers
  \texttt{recvbuf} and \texttt{sendbuf} are supplied by the
  \isoneighboralltoall call, while \texttt{interbuf} is an
  intermediate buffer of the same size. The $\mathrm{firstround}[i]$
  for neighbor $i$ is the first dimension $j$ with $c_j^i\neq 0$. The
  RECV and SEND output consisting of $k$ parts represents the
  schedule, and describes what happens for neighbor block $i$ in
  communication step $h$. RECV and SEND is used to set up the MPI derived
  datatypes for immediate neighbor communication in each step. Only
  the computations for positive coordinates are shown.}
\label{alg:zero-alltoall}
\end{algorithm}

So far we did not describe how the blocks to send and receive are
combined in the steps of the communication rounds. We now present the
full schedule computation for the \alltoall operation. In each of the
$D$ communication steps (see Proposition~\ref{prop:alltoall}), at
least one new block is received and one block one sent. The initial
blocks are present in the send buffer given in the
\isoneighboralltoall call (which must not be changed), and eventually
all source blocks have to be received into the given receive
buffer. Over the communication rounds, the block to the $i$th neighbor
$R\oplus C^i$ will traverse $\|C^i\|$ hops. We will let the block
alternate between intermediate and receive buffers of the processes
that it traverses, such that it ends up in the $i$th position of the
receive buffer at process $R\oplus C^i$ in the last round.  In each
communication step, some blocks are sent from the intermediate buffer
and received into the receive buffer, and other blocks are sent from
the receive buffer and received into the intermediate buffer.  A block
will end up in the receive buffer if we receive it into the receive
buffer when there are an odd number of hops is remaining.  In each
step of the schedule, all blocks to be sent in that step are combined
into one message; likewise for the blocks received. Instead of doing
this explicitly by copying into yet another intermediate buffer, two
MPI derived datatypes are constructed, one describing the blocks to be
received (whether into receive or intermediate buffer) and one
describing the blocks to be sent. These MPI datatypes consist of $k$
parts corresponding to the $k$ blocks sent and received in that step.
Since blocks are located in one of three different buffers (send,
receive and intermediate), an MPI structured type is needed and
constructed with \mpistruct. With these derived datatypes, the MPI
send and receive operations directly access the blocks from the
corresponding buffers without any need for explicit packing and
unpacking from contiguous communication buffers. The same kind of
block-by-block double buffering with derived datatypes was used by
Tr\"aff~\etal~\cite{Traff14:bruck}. This is our final, so-called
\emph{zero-copy implementation}: all data movement operations between
buffers are done implicitly by MPI communication operations using the
MPI derived datatypes constructed by the schedule without any
process-local, explicit copying of blocks to be sent or received. The
construction of the alternating buffer schedule is shown as
Algorithm~\ref{alg:zero-alltoall}. The schedule is represented by the
$D$ send and receive datatypes for the $D$ communication steps. The
schedule is precomputed at the \isoneighboralltoallinit call, so that
data type creation can be amortized over the ensuing \isostart
calls.

\section{Experimental Evaluation, Part One}
\label{sec:experiment1}

In order to assess the potential gains of zero-copy message-combining,
we compare our isomorphic collective implementations to the MPI neighborhood
collectives that express the same communication patterns, namely
\mpineighboralltoall, \mpineighborallgather, and
\mpineighboralltoallw.

For our basic comparisons, we use generalizations of the
application-relevant, two-dimensional $9$-point stencil
pattern, so-called \emph{\Moore neighborhoods}\footnote{See
  \url{http://mathworld.wolfram.com/MooreNeighborhood.html}, last
  visited on April 8, 2016.}~\cite{ToffoliMargolus87}. A
$d$-dimensional, radius $r$ Moore neighborhood consists of all
neighbors $C^i$ whose largest absolute value coordinate $|c^i_j|$ is
at most $r$.
\Moore neighborhoods have large numbers of neighbors, namely
$s=(2r+1)^{d}-1$ (excluding the process itself), which can reduce the
number of communication rounds from $s$ down to $D=2rd$ for the
torus-based message-combining algorithms. 
\begin{table}[t]
  \centering
  \caption{Parallel machines used in experiments.}
  \label{tab:machines}
  \begin{footnotesize}
  \begin{tabular}{l@{\hskip .05in}l@{\hskip .05in}l}
    \toprule
    name & hardware & MPI libraries \\
    \midrule 
    \machjupiter & \num{36} Dual Opteron 6134\,@\,\SI{2.3}{\giga\hertz}  & \jupiternecmpi \\
             &  \infiniband QDR MT4036  &  \\
    \machvsc &  \num{2000} Dual Xeon E5-2650V2\,@\,\SI{2.6}{\giga\hertz} & \vscmvapich  \\
             & \infiniband QDR-80 &  \\
    \macharcher &  \num{4920} Dual Xeon E5-2697V2\,@\,\SI{2.7}{\giga\hertz} & \archermpi  \\ 
             &  Cray Dragonfly &  \\
    \bottomrule
  \end{tabular}
  \end{footnotesize}
\end{table}
Initial experiments were conducted on a small \num{36}~node cluster.
We expect the performance to depend mostly on the neighborhood, and
less on the number of processes.  To corroborate, we repeated the
experiments on \num{70}~and \num{500}~nodes of two larger systems using 
different MPI libraries. The system configurations are summarized in
\tab~\ref{tab:machines}.

In each experiment we measure the \runtime of either \alltoall or
\allgather implementations over different, small block sizes.  We
perform \num{100}~repetitions of each measurement and synchronize MPI
processes before each measurement.  We compute the \runtime by taking
the maximum local \runtime across all processes in the collective
operation. Each experiment is repeated \num{10}~times to account for
\runtime variations across individual \mpiruns.  Processes are always
pinned to specific cores, and the CPU frequency is set as high as
possible.
We remove outliers with Tukey's method (using a bound of three times
the inter-quartile range), and we compute the median \runtime of the
remaining measurements. Results are shown as bar plots of the median
of the previously obtained medians over the \num{10}~\mpiruns, along
with their minimum and maximum values to visualize possible \runtime
variations.

\begin{table}
\caption{\Setup times of neighborhoods and schedule computations 
(median of \num{400}~measurements, \Moore neighborhood,  
\num{30x16} processes, \jupiternecmpi, \machjupiter).}
\label{tab:setuptimes}
{\footnotesize
\begin{tabular}{llrrr}
  \toprule
 & &\multicolumn{3}{c}{Radius} \\
 \cmidrule(l){3-5}\#Dims & Function & 1 & 2 & 3 \\ 
  & & [ms] & [ms] & [ms] \\
\midrule\multirow{4}{*}{\textbf{2}} & MPI\_Dist\_graph\_create\_adjacent & 0.20 & 0.14 & 0.16 \\ 
   & MPI\_Dist\_graph\_create & 27.00 & 95.78 & 158.38 \\ 
   & Iso\_neighborhood\_create & 0.02 & 0.01 & 0.02 \\ 
   & Iso\_neighborhood\_alltoall\_init & 0.10 & 0.05 & 0.06 \\ 
  \midrule\multirow{4}{*}{\textbf{3}} & MPI\_Dist\_graph\_create\_adjacent & 0.13 & 0.16 & 0.23 \\ 
   & MPI\_Dist\_graph\_create & 130.75 & 489.86 & 939.38 \\ 
   & Iso\_neighborhood\_create & 0.02 & 0.05 & 0.14 \\ 
   & Iso\_neighborhood\_alltoall\_init & 0.03 & 0.10 & 0.37 \\ 
  \midrule\multirow{4}{*}{\textbf{4}} & MPI\_Dist\_graph\_create\_adjacent & 0.17 & 0.37 & 1.04 \\ 
   & MPI\_Dist\_graph\_create & 6.17 & 233.13 & 337.32 \\ 
   & Iso\_neighborhood\_create & 0.04 & 0.28 & 1.06 \\ 
   & Iso\_neighborhood\_alltoall\_init & 0.06 & 0.42 & 2.94 \\ 
  \midrule\multirow{4}{*}{\textbf{5}} & MPI\_Dist\_graph\_create\_adjacent & 0.25 & 1.48 & 7.34 \\ 
   & MPI\_Dist\_graph\_create & 7.42 & 28.65 & 42.22 \\ 
   & Iso\_neighborhood\_create & 0.11 & 1.48 & 8.13 \\ 
   & Iso\_neighborhood\_alltoall\_init & 0.15 & 2.82 & 25.02 \\ 
   \bottomrule
\end{tabular}
}

\vspace{-5pt}
\end{table}

\begin{figure*}[t]
\centering
\begin{subfigure}[$d=2$ dimensions, radius $r=1$ (\num{8}~neighbors).]%
{%
        \begingroup
        \input{#{figs/fig2}}%
        \endgroup
    
\label{exp:dim2rad1-moore}%
}%
\end{subfigure}%
\hfill%
\begin{subfigure}[$d=3$ dimensions, radius $r=1$ (\num{26}~neighbors).]%
{%
        \begingroup
        \input{#{figs/fig3}}%
        \endgroup
    
\label{exp:dim3rad1-moore}%
}
\end{subfigure}
\begin{subfigure}[$d=4$ dimensions, radius $r=1$ (\num{80}~neighbors).]%
{%
        \begingroup
        \input{#{figs/fig4}}%
        \endgroup
    
\label{exp:dim4rad1-moore}%
}
\end{subfigure}
\hfill%
\begin{subfigure}[$d=5$ dimensions, radius $r=1$ (\num{242}~neighbors).]%
{%
        \begingroup
        \input{#{figs/fig5}}%
        \endgroup
    
\label{exp:dim5rad1-moore}%
}
\end{subfigure}
\begin{subfigure}[$d=3$ dimensions, radius $r=3$ (\num{342}~neighbors).]%
{%
        \begingroup
        \input{#{figs/fig6}}%
        \endgroup
    
\label{exp:dim3rad3-moore}%
}
\end{subfigure}
\hfill%
\begin{subfigure}[Asymmetric neighborhood (positive coordinates), $d=3$ dimensions, radius $r=3$ (\num{63}~neighbors).]%
{%
        \begingroup
        \input{#{figs/fig7}}%
        \endgroup
    
\label{exp:dim3rad3-asymmoore}%
}
\end{subfigure}
\caption{\label{exp:alltoall} Median \runtimes of \isoneighboralltoall and
 \mpineighboralltoall,  \Moore neighborhood, row order of neighbors,  \num{30x16}~processes, 
 \jupiternecmpi, machine: \machjupiter.}
\end{figure*}

Our first set of experiments compares our message-combining \alltoall
algorithms to the \mpineighboralltoall collective on a series of
\Moore neighborhoods. This is a regular exchange operation, and all
blocks have the same size. The measured \runtimes are shown for
different block sizes. Neighborhoods for the MPI collectives have to
be set up using one of the two distributed graph constructors
\mpidistgraphcreate or \mpidistgraphcreateadjacent, which can both be
rather costly. In Table~\ref{tab:setuptimes} we compare the \setup
times for the full \Moore neighborhoods used in the experiments for
dimension $d=2, 3, 4, 5$ and radius $r=1, 2, 3$. As expected, the
\mpidistgraphcreate constructor is significantly more expensive than
the more specific \mpidistgraphcreateadjacent, with an unexplained
drop in the MPI \setup times when going from \num{3} to \num{4}
dimensions.  Our \isoneighborhoodcreate is faster than or at least in
the same ballpark as \mpidistgraphcreateadjacent.  We also report the
time for \isoneighboralltoallinit, in which the schedule computation
of \alg~\ref{alg:zero-alltoall} is performed, including the creation
of the MPI derived datatypes. With our interface, 
setup and initialization time can be amortized over several
\isoneighboralltoall calls, still it is important that these times be as
low as possible.

For the underlying Cartesian communicator of the isomorphic
neighborhoods, we use \mpidimscreate (despite its potential
problems~\cite{Traff15:dimscreate}) and enable reordering, such that
the virtual torus may be aligned with the underlying communication
system.

For higher dimensions of the tested neighborhoods, the number of
relative neighbors is larger than the number of processes, such that
the same process is a neighbor for many different blocks. Our
implementations work regardless, and all such block are combined into
the same message.

Our communication experiments use small block sizes from \SI{1}{\byte}
to \SI{2}{\kilo\byte}.  Selected results for \Moore neighborhoods in
dimension $d=2,3,4,5$ with radius $r=1,3$ are shown in
\fig~\ref{exp:dim2rad1-moore} to \fig~\ref{exp:dim3rad3-moore}. For
small block sizes, we observe considerable improvements over the MPI
neighborhood collectives, close to the ratio of number of neighbors to
$2d$.  It is interesting to note that the performance of the MPI
neighborhood collectives sometimes depends on whether the neighborhood
was set up with \mpidistgraphcreate or \mpidistgraphcreateadjacent. As
block sizes grow, the advantage of message-combining diminishes.
Finally, the experiment in \fig~\ref{exp:dim3rad3-asymmoore} considers
isomorphic \alltoall communication with asymmetric neighborhoods, and
shows the benefits of zero-copy message-combining in this situation.
We used an incomplete \Moore neighborhood in $d=3$ dimensions and
radius $r=3$ consisting only of the positive coordinate neighbors, as
in Section~\ref{sec:definitions}.

\begin{figure*}[t]
\centering
\begin{subfigure}[$d=3$ dimensions, radius $r=1$ (\num{26}~neighbors).]%
{%
        \begingroup
        \input{#{figs/fig8}}%
        \endgroup
    
\label{exp:alltoallw-irregular-d3r1}%
}%
\end{subfigure}%
\hfill%
\begin{subfigure}[$d=4$ dimensions, radius $r=1$ (\num{80}~neighbors).]%
{%
        \begingroup
        \input{#{figs/fig9}}%
        \endgroup
    
\label{exp:alltoallw-irregular-d4r1}%
}
\end{subfigure}
\caption{\label{exp:alltoallw-irregular} Median \runtimes of \isoneighboralltoallw and \mpineighboralltoallw, \Moore neighborhood with irregular data distribution to neighbors, 
row order of neighbors,  \num{30x16}~processes, \jupiternecmpi, machine: \machjupiter.}
\end{figure*}

Our implementation of the irregular \isoneighboralltoallw operation,
which uses the same schedules as in the regular case, is benchmarked
in \fig~\ref{exp:alltoallw-irregular}.  The plots show the results of
the experiment with an irregular data distribution.  Here, the block
sizes sent to each neighbor depend on the distance of that neighbor
$\|C^i\|$, such that the block sent to neighbor $i$ is of size
$\hat{m}^{d-\|C^i\|}$. This emulates the behavior of many stencil
computations, where the messages exchanged with corners are smaller
than with edges and hyperplanes.  We tested the algorithm with three-
and four-dimensional \Moore neighborhoods with radius $r=1$, having
\num{26}~and \num{80}~neighbors, respectively.  The base block size
$\hat{m}$ is varied between the different experiments and is shown on
the x-axis, together with the total size of the send buffer per
process.  For example, in \fig~\ref{exp:alltoallw-irregular-d3r1}, for
$\hat{m}=\SI{512}{\byte}$, each process sends messages with one of the
following sizes to the \num{26} neighbors: \SI{1}{\byte},
\SI[exponent-base = 512]{e1}{\byte}, and \SI[exponent-base =
  512]{e2}{\byte}, amounting to a total size of \SI{1.5}{\mega\byte}
for the entire send buffer.  In the experiment
(see \fig~\ref{exp:alltoallw-irregular}), our \alltoallw implementation
outperforms the standard MPI collective in most of the cases.

\begin{figure*}[t]
\centering
\begin{subfigure}[\Moore neighborhood in $d=3$ dimensions, radius $r=3$ (\num{342}~neighbors).]%
{%
        \begingroup
        \input{#{figs/fig10}}%
        \endgroup
    
\label{exp:dim3rad3-direct}%
}%
\end{subfigure}%
\hfill%
\begin{subfigure}[``Shales'' corresponding to radius $r_1=3,r_2=7$ in a \Moore neighborhood in $d=3$ dimensions (\num{1396}~neighbors).]%
{%
        \begingroup
        \input{#{figs/fig11}}%
        \endgroup
    
\label{exp:dim3shales-direct}%
}
\end{subfigure}
\caption{\label{exp:alltoall-pers-vs-direct} Median \runtimes of the straightforward neighbor 
\alltoall implementation, \isoneighboralltoall and \isoneighboralltoalldirect, row order of neighbors, 
\num{30x16} processes, \jupiternecmpi, machine: \machjupiter.}
\end{figure*}

\begin{figure*}[t]
\centering
\begin{subfigure}[$d=3$ dimensions, radius $r=3$ (\num{342}~neighbors).]%
{%
        \begingroup
        \input{#{figs/fig12}}%
        \endgroup
    
\label{exp:dim3rad3-allgathermoore}%
}
\end{subfigure}
\hfill%
\begin{subfigure}[Asymmetric (positive coordinates) neighborhood in $d=3$ dimensions, radius $r=3$ (63 neighbors).]%
{%
        \begingroup
        \input{#{figs/fig13}}%
        \endgroup
    
\label{exp:dim3rad3-allgatherasym}%
}%
\end{subfigure}%
\caption{\label{exp:allgather} Median \runtimes of \isoneighborallgather, \isoneighboralltoall and \mpineighborallgather, 
\Moore neighborhood, row order of neighbors, \num{30x16}~processes, \jupiternecmpi, machine: \machjupiter.}
\end{figure*}

\section{Better Algorithms}
\label{sec:optimizations}

The assumption of a one-ported torus network was useful in that it led
to easily computable, optimal message-combining schedules. However,
most real systems (e.g., as in \tab~\ref{tab:machines}) have 
different, more powerful communication systems. If we relieve the torus
assumption, better algorithms for more powerful communication systems
may be possible, and interesting optimization problems and tradeoffs
between the number of communication rounds and volume
arise~\cite{Bruck97}.

Assume that we have---at the other extreme---a fully connected,
bidirectional, $k$-ported communication system. In this case, we could
ask: What is the minimal number of communication rounds for a given
$s$-neighborhood? What is the optimal load balance in number of blocks
sent per communication round? What is the optimal schedule for an
irregular $s$-neighborhood where blocks to be sent to different
neighbors may have different sizes?

To minimize the number of communication rounds in a one-ported,
fully-connected system, the following optimization problem has to be
solved.  Given a set of $s$ vectors $\cal C$, find a smallest
\emph{additive basis} $\cal B$ such that each $C\in\cal C$ can be
written as a sum of distinct $B_i\in \cal B$. Note that it is
explicitly not required that ${\cal B}\subseteq {\cal C}$. Our torus
algorithms use the additive basis vectors
$(1,0,0,\ldots),(0,1,0,\ldots),(0,0,1,\ldots)$, but in general need
repetitions (several hops) of the basis vectors. The algorithm that
will be sketched below uses distinct basis vectors. Given an additive
basis, we claim that a schedule can be computed easily and similarly
to the torus schedules, and both \alltoall and \allgather operations
will require $|\cal B|$ rounds. How hard is the problem of finding
smallest additive bases for arbitrary $s$-neighborhoods?  Some $d=1$
dimensional examples are illustrative. For ${\cal C}=\{1,2,3\}$, a
minimal additive basis is $\{1,2\}$.  For ${\cal
  C}=\{1,2,3,4,5,6,7\}$, a minimal additive basis is $\{1,2,4\}$,
which is the scheme used by logarithmic doubling \alltoall and
\allgather algorithms~\cite{Bruck97}.  For ${\cal
  C}=\{1,2,3,4,5,6,7,8\}$, minimal additive bases are
$\{1,2,3,6\}$ or $\{1,2,4,8\}$.

Let us assume instead a $d$-dimensional torus communication system
with direct communication along the dimensions, such that it is
possible to send a message directly to a neighbor with relative
coordinate $c_j$ in any of the dimensions. We can perform the
communication operations using an additive (but not necessarily
minimal) basis consisting of all projected vectors
$(0,\ldots,c_j,\ldots,0)$ for the different $c_j$ in each of the $d$
dimensions. We can easily modify our schedules to use this basis,
namely to send directly to relative neighbor $(0,\ldots,c_j,\ldots,0)$
instead of via $c_j$ hops. All blocks going to the same relative
neighbor in round $j$ can be combined. In order to achieve this, in
communication round $j$ the relative neighbors need to be (bucket)
sorted for the $j$th dimension. For each neighbor, the number of hops
to traverse is reduced from $\|C^i\|$ to the number of non-zero
coordinates in $C^i$, and summing this over all $s$ neighbors gives
the total number of messages sent. The number of rounds needed per
dimension is the number of different, non-zero coordinates, and
summing over all dimensions gives the total number of rounds.  Since
the number of rounds is no longer dependent on the magnitude of the
coordinates, schedules can now be computed in $O(sd)$ operations.

We have implemented both \isoneighboralltoall and
\isoneighborallgather along these lines which we call the \emph{torus
  direct}
algorithms. For non-torus systems, e.g., those in
Table~\ref{tab:machines}, we expect that direct communication can be
exploited so that the smaller number of communication rounds
will indeed pay off.

\section{Experimental Evaluation, Part Two}
\label{sec:experiment2}

\begin{figure*}[t]
\centering
\begin{subfigure}[\num{70x1}~processes.]%
{%
        \begingroup
        \input{#{figs/fig14}}%
        \endgroup
    
\label{exp:vsc3-nnp1}%
}%
\end{subfigure}%
\hfill%
\begin{subfigure}[\num{70x16}~processes.]%
{%
        \begingroup
        \input{#{figs/fig15}}%
        \endgroup
    
\label{exp:vsc3-nnp16}%
}
\end{subfigure}
\caption{\label{exp:vsc-alltoall-pers-vs-direct} Median \runtimes of \isoneighboralltoall and 
\isoneighboralltoalldirect (neighborhood set up using \isoneighborhoodcreate), 
\Moore neighborhood in $d=3$ dimensions, radius $r=3$ (\num{342}~neighbors), 
row order of neighbors, \vscmvapich, machine: \machvsc.}
\end{figure*}

\begin{figure}[t]
\centering
        \begingroup
        \input{#{figs/fig16}}%
        \endgroup
    
\caption{\label{exp:vsc3-comparison} Median \runtimes of \isoneighboralltoall and 
\isoneighboralltoalldirect, \Moore neighborhood, row order of neighbors, 
\num{70x16}~processes, \vscmvapich, machine: \machvsc.}
\end{figure}

\begin{figure*}[t]
\centering
\begin{subfigure}[$d=4$ dimensions, radius $r=1$ (\num{80}~neighbors).]%
{%
        \begingroup
        \input{#{figs/fig17}}%
        \endgroup
    
\label{exp:archer-alltoall-pers-vs-trivial-d4}%
}%
\end{subfigure}%
\hfill%
\begin{subfigure}[$d=5$ dimensions, radius $r=1$ (\num{242}~neighbors).]%
{%
        \begingroup
        \input{#{figs/fig18}}%
        \endgroup
    
\label{exp:archer-alltoall-pers-vs-trivial-d5}%
}
\end{subfigure}
\caption{\label{exp:archer-alltoall-pers-vs-trivial} Median \runtimes of \isoneighboralltoall,  
\mpineighboralltoall and the straightforward implementation of \isoneighboralltoall, \Moore neighborhood, row order of neighbors, 
\num{500x24}~processes, \archermpi, machine: \macharcher.}
\end{figure*}

We have benchmarked the torus direct implementations using the
same systems and \Moore neighborhoods as in
\Sec~\ref{sec:experiment1}, but the emphasis is on comparing our three
implementations, namely the straightforward 
implementation shown in \lst~\ref{lst:s-round}, the optimal torus
implementations, and the torus direct algorithms.  Selected
results for \isoneighboralltoall are shown in
\fig~\ref{exp:dim3rad3-direct}, while in
\fig~\ref{exp:dim3shales-direct} we have used a neighborhood
consisting of ``shales'' of neighbors at the Chebyshev distances $r_1=3$
and $r_2=7$. As message sizes grow, the smaller number of
communication rounds and the smaller total communication volume of the
torus direct algorithm make it perform gradually better than the
optimal torus algorithm. For the shales neighborhood in
\fig~\ref{exp:dim3shales-direct}, the number of communication rounds
for the torus algorithm is about $2 r_2 d =42$ compared to only $(2+2)
d=12$ for the direct algorithm, and, more significantly, in the former
the number of times blocks are sent further on is proportional to the
number of rounds. The torus algorithm becomes slower than the
straightforward algorithm already for message sizes of
\SI{500}{\byte}; in contrast, the direct algorithm stays on
par with the straightforward one in the message range shown. The
experiments show that exploiting direct communication can lead to
better performing message-combining implementations; it is therefore
relevant to pursue the optimization problems posed in
Section~\ref{sec:optimizations}.

The \isoneighborallgather collective is investigated in
\fig~\ref{exp:dim3rad3-allgathermoore} with a complete
three-dimensional \Moore neighborhood and in
\fig~\ref{exp:dim3rad3-allgatherasym} with an asymmetric \Moore
neighborhood. The \runtimes of the \mpineighborallgather operation are
similar to those of the \mpineighboralltoall for the same
neighborhood, as can be seen for small message sizes by comparing to
\fig~\ref{exp:dim3rad3-moore} and \fig~\ref{exp:dim3rad3-asymmoore}.
Thus, \fig~\ref{exp:allgather} suggests that the MPI library we used
implements the \allgather and \alltoall operations in exactly the same
way: each block of data is sent directly to the corresponding
neighbor. In contrast, the \isoneighborallgather operation achieves an
\SI{80}{\percent} \runtime reduction over \mpineighborallgather for
the tested message sizes, as well as a substantially improved
performance over its \alltoall counterpart.  This behavior can be
explained by the design of the \allgather schedule, which reduces the
volume of data sent, compared to the \alltoall one.  To further
highlight the efficiency of the \allgather schedule, we compare
\isoneighborallgather with \isoneighboralltoall for an asymmetric
Moore neighborhood in \fig~\ref{exp:dim3rad3-allgatherasym}. Here, we
can again see that using \isoneighborallgather pays off as the message
size increases, as the operation completes three times faster than
\alltoall for message sizes of up to \SI{40}{\kilo\byte}.

Finally, we have evaluated the proposed torus implementations on the
\machvsc machine, using the \vscmvapich~library, and the \macharcher
machine with the \archermpi~library. As in this scenario we do not
have dedicated access to the entire machine, we have conducted
\num{300}~measurements for each collective operation to compensate for
the possible variations and we have repeated each experiment
\num{10}~times.  \fig~\ref{exp:vsc-alltoall-pers-vs-direct}  compares
the \runtimes of the optimal torus \alltoall and the torus direct
algorithm with the MPI neighborhood \alltoall implementation and the
straightforward algorithm shown in
Listing~\ref{lst:s-round}. Similarly to our previous experiments, this
scenario emphasizes the advantage of the direct strategy in the case
of a fully-connected hardware topology. While \isoneighboralltoall
outperforms the MPI implementation only for smaller message sizes, the
direct algorithm achieves the best \runtime performance up to
\SI{1}{\kilo\byte}.  For message sizes under \SI{512}{\byte}, both
implementations outperform the straightforward algorithm in the
\num{70x1}~processes scenario.  When the total data size exchanged
increases in \fig~\ref{exp:vsc3-nnp16}, our implementations show less
improvement due to the larger number of processes per node.

\fig~\ref{exp:vsc3-comparison} compares the torus \alltoall
implementations with the straightforward algorithm. Even though the
neighborhood size is comparable to the first scenario, the overhead of
the optimal torus \alltoall algorithm relative to the direct algorithm
is smaller, showing the impact of the size of the neighborhood radius
(and therefore of the number of hops along each dimension) on the
operation \runtime. Nevertheless, for small message sizes both
implementations provide better results than the straightforward
\alltoall algorithm. 

\fig~\ref{exp:archer-alltoall-pers-vs-trivial} shows our results on
\macharcher. The MPI collectives perform much better here than was the
case for the other machines, such that our message-combining algorithms
for the small $r=1$ case show only little advantage. The MPI
neighborhood collectives can apparently use the pipelining and
multi-ported capabilities of the \macharcher network better than our
send and receive based implementations. We have therefore compared our
message combining algorithm with the straightforward algorithm of
Listing~\ref{lst:s-round}, over which we can improve by large factors
(as for the other machines). Again, this shows that finding additive
bases that allow for many simultaneous communication operations is an
important optimization problem (\Sec~\ref{sec:optimizations}).

\section{Summary}
\label{sec:summary}

We proposed a specification for isomorphic (sparse) collective
communication to derive simple, message-combining algorithms for
\alltoall and \allgather type of sparse collective communication
operations. We outlined two types of algorithms, one assuming a torus
communication network that is optimal in both the number of
communication rounds and the total number of messages sent, and one
assuming a more liberal torus allowing direct communication along the
torus dimensions that reduces both the number of rounds and the
communication volume. The latter algorithm is an in-between the torus
algorithm and an algorithm using direct communication between
neighbors. Both types of algorithms were implemented and compared to
typical implementations of the corresponding MPI neighborhood
collective communication operations, against which our implementations
perform significantly better for smaller message sizes.  In our
experiments we used (also asymmetric) variations of the \Moore
neighborhoods. The experiments show that there is large room for
improvements of current implementations of the MPI neighborhood
collectives. Our algorithms could potentially be used to obtain such
improvements, but only if it is externally asserted (or can easily be
detected) that neighborhoods are indeed isomorphic.

Our isomorphic neighborhoods are embedded in $d$-dimensional tori, but
our schedules can easily be extended to non-periodic tori, as can be
defined with MPI Cartesian topologies. Furthermore, it would be
possible to extend the idea of isomorphic neighborhoods also to other
regular underlying virtual topologies. Our experiments were performed
on non-torus systems, for which the virtual torus topology used to
describe relative neighborhoods is only a convenience. It would be
interesting to perform experiments on actual torus systems (Blue Gene
or K Computer), where the virtual topology has actually been mapped
efficiently onto the hardware topology.

For stencil-type computations, non-blocking communication is natural
to potentially overlap parts of the stencil update with neighborhood
communication. The proposed, persistent interface has a blocking
\isostart operation. Similarly to what is currently being discussed in
the MPI community, it could be declared non-blocking by adding the
following call

\begin{lstlisting}
Iso_wait(Iso_request *request);
\end{lstlisting}

\noindent
at which local completion can be enforced. We think that this is a
valuable extension, for which algorithms and implementations should be
developed.

\balance
\bibliographystyle{IEEEtran}
\bibliography{traff,parallel}

\begin{thebibliography}{10}
\providecommand{\url}[1]{#1}
\csname url@samestyle\endcsname
\providecommand{\newblock}{\relax}
\providecommand{\bibinfo}[2]{#2}
\providecommand{\BIBentrySTDinterwordspacing}{\spaceskip=0pt\relax}
\providecommand{\BIBentryALTinterwordstretchfactor}{4}
\providecommand{\BIBentryALTinterwordspacing}{\spaceskip=\fontdimen2\font plus
\BIBentryALTinterwordstretchfactor\fontdimen3\font minus
  \fontdimen4\font\relax}
\providecommand{\BIBforeignlanguage}[2]{{%
\expandafter\ifx\csname l@#1\endcsname\relax
\typeout{** WARNING: IEEEtran.bst: No hyphenation pattern has been}%
\typeout{** loaded for the language `#1'. Using the pattern for}%
\typeout{** the default language instead.}%
\else
\language=\csname l@#1\endcsname
\fi
#2}}
\providecommand{\BIBdecl}{\relax}
\BIBdecl

\bibitem{Sourcebook03}
J.~Dongarra, I.~Foster, G.~Fox, W.~Gropp, K.~Kennedy, L.~Torczon, and A.~White,
  Eds., \emph{Sourcebook of Parallel Computing}.\hskip 1em plus 0.5em minus
  0.4em\relax Morgan Kaufmann Publishers, 2003.

\bibitem{Epperson07}
J.~F. Epperson, \emph{An Introduction to Numerical Methods and Analysis},
  2nd~ed.\hskip 1em plus 0.5em minus 0.4em\relax Wiley-Interscience, 2013.

\bibitem{CuiOlsen10}
Y.~Cui, K.~B. Olsen, T.~H. Jordan, K.~Lee, J.~Zhou, P.~Small, D.~Roten, G.~Ely,
  D.~K. Panda, A.~Chourasia, J.~M. Levesque, S.~M. Day, and P.~Maechling,
  ``Scalable earthquake simulation on petascale supercomputers,'' in
  \emph{Conference on High Performance Computing Networking, Storage and
  Analysis ({SC})}, 2010, pp. 1--20.

\bibitem{MPI-3.1}
{MPI Forum}, \emph{\textsf{MPI}: A Message-Passing Interface Standard. Version
  3.1}, June 4th 2015, \url{www.mpi-forum.org}.

\bibitem{Traff15:isosparse}
J.~L. Tr{\"a}ff, F.~D. L{\"u}bbe, A.~Rougier, and S.~Hunold, ``Isomorphic,
  sparse {MPI}-like collective communication operations for parallel stencil
  computations,'' in \emph{22nd European MPI Users' Group Meeting
  ({EuroMPI})}.\hskip 1em plus 0.5em minus 0.4em\relax ACM, 2015.

\bibitem{Bruck97}
J.~Bruck, C.-T. Ho, S.~Kipnis, E.~Upfal, and D.~Weathersby, ``Efficient
  algorithms for all-to-all communications in multiport message-passing
  systems,'' \emph{{IEEE} Transactions on Parallel and Distributed Systems},
  vol.~8, no.~11, pp. 1143--1156, 1997.

\bibitem{Traff14:bruck}
J.~L. Tr{\"a}ff, A.~Rougier, and S.~Hunold, ``Implementing a classic: Zero-copy
  all-to-all communication with {MPI} datatypes,'' in \emph{28th {ACM}
  International Conference on Supercomputing ({ICS})}.\hskip 1em plus 0.5em
  minus 0.4em\relax ACM, 2014, pp. 135--144.

\bibitem{BasuHallWilliamsStraalenOlikerColella15}
P.~Basu, M.~W. Hall, S.~Williams, B.~van Straalen, L.~Oliker, and P.~Colella,
  ``Compiler-directed transformation for higher-order stencils,'' in \emph{2015
  {IEEE} International Parallel and Distributed Processing Symposium,
  ({IPDPS})}, 2015, pp. 313--323.

\bibitem{Dursun09}
H.~Dursun, K.~Nomura, L.~Peng, R.~Seymour, W.~Wang, R.~K. Kalia, A.~Nakano, and
  P.~Vashishta, ``A multilevel parallelization framework for high-order stencil
  computations,'' in \emph{15th International Euro-Par Conference}, 2009, pp.
  642--653.

\bibitem{Dursun12}
H.~Dursun, M.~Kunaseth, K.~Nomura, J.~Chame, R.~F. Lucas, C.~Chen, M.~W. Hall,
  R.~K. Kalia, A.~Nakano, and P.~Vashishta, ``Hierarchical parallelization and
  optimization of high-order stencil computations on multicore clusters,''
  \emph{The Journal of Supercomputing}, vol.~62, no.~2, pp. 946--966, 2012.

\bibitem{StengelTreibigHagerWellein15}
H.~Stengel, J.~Treibig, G.~Hager, and G.~Wellein, ``Quantifying performance
  bottlenecks of stencil computations using the execution-cache-memory model,''
  in \emph{Proceedings of the 29th {ACM} on International Conference on
  Supercomputing ({ICS})}, 2015, pp. 207--216.

\bibitem{TangChowdhuryKuzmaulLukLeiserson11}
Y.~Tang, R.~A. Chowdhury, B.~C. Kuszmaul, C.~Luk, and C.~E. Leiserson, ``The
  pochoir stencil compiler,'' in \emph{Proceedings of the 23rd Annual {ACM}
  Symposium on Parallelism in Algorithms and Architectures ({SPAA})}, 2011, pp.
  117--128.

\bibitem{BordawekarChoudharyRamanujam96:automatic}
R.~Bordawekar, A.~N. Choudhary, and J.~Ramanujam, ``Automatic optimization of
  communication in compiling out-of-core stencil codes,'' in \emph{Proceedings
  of the 10th international conference on Supercomputing, {ICS}}, 1996, pp.
  366--373.

\bibitem{ZhuZhangYoshiiLiZhangBalaji15}
X.~Zhu, J.~Zhang, K.~Yoshii, S.~Li, Y.~Zhang, and P.~Balaji, ``Analyzing
  {MPI-3.0} process-level shared memory: {A} case study with stencil
  computations,'' in \emph{15th {IEEE/ACM} International Symposium on Cluster,
  Cloud and Grid Computing {(CCGrid)}}, 2015, pp. 1099--1106.

\bibitem{HoeflerSchneider12}
T.~Hoefler and T.~Schneider, ``Optimization principles for collective
  neighborhood communications,'' in \emph{{IEEE/ACM} Conference on High
  Performance Computing Networking, Storage and Analysis ({SC})}, 2012, p.~98.

\bibitem{Ovcharenko12}
A.~Ovcharenko, D.~Ibanez, F.~Delalondre, O.~Sahni, K.~E. Jansen, C.~D.
  Carothers, and M.~S. Shephard, ``Neighborhood communication paradigm to
  increase scalability in large-scale dynamic scientific applications,''
  \emph{{P}arallel {C}omputing}, vol.~38, no.~3, pp. 140--156, 2012.

\bibitem{SouravlasRoumeliotis08:torus}
S.~Souravlas and M.~Roumeliotis, ``A message passing strategy for array
  redistributions in a torus network,'' \emph{The Journal of Supercomputing},
  vol.~46, no.~1, pp. 40--57, 2008.

\bibitem{ToffoliMargolus87}
T.~Toffoli and N.~Margolus, \emph{Cellular Automata Machines: A New Environment
  for Modeling}.\hskip 1em plus 0.5em minus 0.4em\relax {MIT} Press, 1987.

\bibitem{Traff15:dimscreate}
J.~L. Tr{\"a}ff and F.~D. L{\"u}bbe, ``Specification guideline violations by
  \texttt{MPI\_Dims\_create},'' in \emph{22nd European MPI Users' Group Meeting
  ({EuroMPI})}.\hskip 1em plus 0.5em minus 0.4em\relax ACM, 2015.

\end{thebibliography}

\end{document}